 \documentclass[manuscript]{aastex}

\newcommand{\corot}{CoRoT--1b }
\newcommand{\ogle}{OGLE--TR--56b }

\shorttitle{Ks-band thermal emission from \corot}
\shortauthors{Rogers et al.}

\begin{document}


\title{Ks-band detection of thermal emission and color constraints to CoRoT--1b:
A low--albedo planet with inefficient
atmospheric energy redistribution and a 
temperature inversion\altaffilmark{1}
}


\author{Justin C. Rogers\altaffilmark{2}, D\'aniel Apai\altaffilmark{3}, Mercedes L\'opez-Morales\altaffilmark{4}, 
David K. Sing\altaffilmark{5} \& Adam Burrows\altaffilmark{6}
}

\affil{e-mail: rogers@pha.jhu.edu}
\altaffiltext{1}{Based on observations obtained with the Apache Point Observatory 3.5-meter telescope, which is owned and operated by the Astrophysical Research Consortium.}
\altaffiltext{2}{Johns Hopkins University, Department of Physics and Astronomy, 366 Bloomberg Center, 3400 N. Charles Street, Baltimore, MD 21218, USA}
\altaffiltext{3}{Space Telescope Science Institute, 3700 San Martin Drive, Baltimore, MD 21218, USA}
\altaffiltext{4}{Hubble Fellow. Carnegie Institution of Washington, Department of Terrestrial Magnetism, 5241 Broad Branch Rd. NW, Washington D.C., 20015, USA}
\altaffiltext{5}{UPMC Univ Paris 06, CNRS, Institut d'Astrophysique de Paris, 98bis boulevard Arago, F-75014 Paris, France}
\altaffiltext{6}{Princeton University, Department of Astrophysical Sciences, Peyton Hall, Princeton, NJ 08544, USA}

\date{Received \today}


\begin{abstract}
We report the detection in Ks-band of the secondary eclipse of the hot Jupiter 
CoRoT--1b, from time series photometry with the ARC 3.5-m telescope at Apache 
Point Observatory.  The eclipse shows a depth of 0.336$\pm$0.042 percent and 
is centered at phase 0.5022$^{+0.0023}_{-0.0027}$, consistent with a zero 
eccentricity orbit (e $\cos{\omega} = 0.0035^{+0.0036}_{-0.0042}$).  We perform 
the first optical to near--infrared multi--band photometric analysis of an 
exoplanet's atmosphere and constrain the reflected and thermal emissions 
by combining our result with the recent 
0.6, 0.71, and 2.09 $\mu$m secondary eclipse detections by 
\citet{2009Natur.459..543S}, \citet{2009arXiv0905.4571G}, and 
\citet{2009A&A...501L..23A}.  
Comparing the multi-wavelength 
detections to state--of--the--art radiative--convective chemical--equilibrium 
atmosphere models, we find the near--infrared fluxes 
difficult to reproduce.  
The closest blackbody--based and physical models provide the following atmosphere 
parameters: a temperature $T =2454^{+84}_{-170}$ K, a very low 
Bond albedo $A_{B} = 0.000^{+0.087}_{-0.000}$, and an energy redistribution parameter 
$P_n = 0.1$, indicating a small but nonzero amount of heat transfer from the 
day-- to night--side.  
The best physical model suggests 
a thermal inversion layer with an extra optical absorber 
of opacity $\kappa_e=0.05$ cm$^2$ g$^{-1}$, 
placed near the 0.1-bar atmospheric pressure level.  
This inversion layer is located ten times deeper in the atmosphere than the 
absorbers used in models to fit mid-infrared Spitzer
detections of other irradiated hot Jupiters.

\end{abstract}


\keywords{binaries:eclipsing -- planetary systems -- stars:individual 
(CoRoT--1) -- techniques: photometric}




\section{Introduction}\label{sec:intro}

Space--based detections of hot Jupiter atmospheres have flourished 
in recent years.  The Spitzer Space Telescope has successfully detected 
thermal emission from several planets at wavelengths longer than 
3.6 $\mu$m (e.g.~\citealt{2005ApJ...626..523C,2005Natur.434..740D,
2007MNRAS.378..148D,2007Natur.447..183K,2008ApJ...673..526K,
2007Natur.447..691H,2008ApJ...684.1427M,2009arXiv0906.1293M}), 
while near--infrared (1.5--2.5 $\mu$m) observations with the Hubble Space 
Telescope have found evidence for water, carbon monoxide, and carbon dioxide 
in the dayside spectrum of the exoplanet HD 189733b \citep{2009ApJ...690L.114S}.  
These detections have been made during secondary eclipses (when the planets 
pass behind their host stars).  
Important atmospheric absorption signatures have been detected by transit 
observations with these space telescopes as well: sodium in HD 209458b 
\citep{2002ApJ...568..377C} and methane in HD 189733b \citep{2008Natur.452..329S} 
with Hubble, and water vapor in HD 189733b with Spitzer \citep{2007Natur.448..169T}.

The CoRoT mission recently joined these space--borne successes by 
detecting phase brightness variations and combined thermal and reflected
emission during secondary eclipses of the exoplanets \corot 
\citep{2009Natur.459..543S,2009A&A...501L..23A} and CoRoT--2b 
\citep{2009arXiv0906.2814A} in an optical broadband window centered 
at about 0.6 $\mu$m (see Table~\ref{table1}).

At the same time that these detections have provided very valuable insights into 
the atmospheric physics of irradiated hot Jupiters, they have also revealed some 
perplexing findings.  
Some of the planets show a strong temperature contrast between their day and night 
sides (e.g.~\citealt{2006Sci...314..623H,2009Natur.459..543S}), while others appear 
to have a more efficient redistribution of incident energy 
(e.g.~\citealt{2007Natur.447..183K}).  
Another striking discovery is an apparent bifurcation of hot Jupiters into 
two classes based on the presence or absence of a thermal inversion layer 
(e.g.~\citealt{2003ApJ...594.1011H,2008ApJ...682.1277B,2008ApJ...678.1419F}).  
Along with a wider--than--expected range of exoplanet radii (see 
e.g.~\citealt{2008A&A...482L..17B}), these are currently the key unsolved 
questions in the study of irradiated hot Jupiter atmospheres.

Resolving these questions will require observations from the optical to 
the infrared wavelength regimes.  
Ground-based observations have just recently started to reach the sensitivity 
necessary to directly detect hot Jupiters.  
The first two detections were announced in early 2009 at optical and 
near--infrared wavelengths:  \ogle in z'-band \citep{2009A&A...493L..31S}, 
and TrES-3 in K-band \citep{2009A&A...493L..35D}.  
The third ground--based result was the detection of an eclipse of \corot 
at 2.09 $\mu$m \citep{2009arXiv0905.4571G}.  
This last detection makes \corot the first exoplanet to have thermal 
emission measured at both optical and near--infrared wavelengths.

Here we report the fourth ground-based detection of thermal emission from 
an exoplanet, and the fourth detection of CoRoT--1b, this time in 
Ks-band (2.15 $\mu$m).  
We combine our results with the other three recent detections of 
\corot by \citet{2009Natur.459..543S}, \citet{2009arXiv0905.4571G}, and 
\citet{2009A&A...501L..23A} to produce the first multi--color analysis 
of the atmospheric spectrum of an exoplanet between 0.5 and 2.2 $\mu$m.

Section \ref{sec:obs} describes the observations.  In section \ref{sec:raa} 
we detail our reduction and analysis steps to detect the signal from the planet.  
Section \ref{sec:models} compares the detected signals to the predictions made by 
state-of-the-art planetary atmosphere models. The results are then discussed 
and summarized in sections \ref{sec:discussion} and \ref{sec:summary}.

\section{Observations}\label{sec:obs}

We observed two secondary eclipse events of \corot on the nights of 2009 January 9 
and 15, UT, using the Near-Infrared Camera \& Fabry-Perot Spectrometer (NICFPS) 
on the ARC 3.5-m telescope at Apache Point Observatory in New Mexico.  
Windy conditions and thin and constantly varying cirrus layers hampered our 
attempts to reach high--precision photometry in the January 9 observations.  
The conditions on January 15 produced good quality data, which we will discuss 
here.

NICFPS is equipped with a Rockwell Hawaii 1-RG 1k$\times$1k HgCdTe detector with a 
4.58$\times$4.58 arc minute field of view and a pixel scale of 0.273 arcsec/pix.  
We chose the reddest available broadband filter (Ks), in order to maximize the 
eclipse depth and collect a large number of photons in as short an exposure 
time as possible.  
The instrument has a high read noise (95 e$^{-}$/pixel), which can be reduced 
through the use of up to 15 Fowler samples.  
In this mode, the chip is read out non--destructively a 
number of times during the exposure, and the difference of each pair 
of readouts is taken as one sample.

We used the instrument in the standard 8 Fowler sample mode, in which 
the chip is read out 16 times consecutively while exposing.  
The first readout is subtracted from the ninth, the second from the tenth, and 
so forth, producing eight 5.44-second samples in 10.88 seconds.  
We then use the average of the eight resulting differences, 
which reduces the read noise of the detector to 35 e$^{-}$/pixel 
(a factor of $\sqrt{8}$).  
To achieve a good sky subtraction, we used a simple two--point dither pattern, 
taking a set of two exposures and then offsetting the objects by 58 pixels 
(15.8 arcsec).  
This resulted in two exposures every $\sim$50 seconds, when combining the 
exposure, readout, and offset times.  
With a measured gain of 4.77 e$^-$/ADU, we collected $\sim$782,000 photons per 
datapoint from the target, and 746,000 to 1,417,000 photons per datapoint from the 
nearby comparison stars.  
Based on the noise expression derived by \citet{1989PASP..101..616H} to combine the 
Poisson noise with the sky level, dark current, and read noise, 
we expected the noise per datapoint to be 0.259\%.

In order to capture the full secondary eclipse and a good baseline for the 
light curve, we monitored the target from 04:20 to 09:40 UT (twice the 
eclipse's duration), obtaining 758 frames over the course of 5.3 hours.  
While the transparency was excellent, the seeing varied between 0.97 to 
1.91 arcsec.  
Throughout the observations, slow drifts in telecope pointing led to a 
varying target location within a $\sim$4-pixel circle.

In addition to the target observations, we took 120 dome flats with dim 
quartz lamps in the Ks filter, and 400 dark frames, each for a range of 
exposure times from 1 to 8 seconds.

\section{Reduction and Analysis}\label{sec:raa}

We began the data reduction by subtracting from each dome flat the average 
of the dark frames for the flat's exposure time, and then combining all the 
flats to create one master normalized flat.  
We then applied the flat-field correction to the target images, and created 
reduced, sky--subtracted images by subtracting from each image the nearest 
neighbor frame taken at the opposite dither position.

Two team members then performed independent analyses of the data, as discussed 
in sections \ref{subsec:raa_a} and \ref{subsec:raa_b}, in order to minimize 
systematics and achieve the optimal photometry.  
In both analyses we converted the JD in the headers of the images to HJD and 
then to an orbital phase using the most recent ephemerides for transits of 
\corot by \citet{2009arXiv0903.1845B}.

\subsection{Analysis A}\label{subsec:raa_a}

The first approach analyzed the combined dither positions as a single dataset.  
Several stars, including the target, were isolated enough in the frames to be 
analyzed by standard aperture photometry.  

Using the IDL adaptation of DAOPHOT, we recorded the flux from CoRoT--1, plus 
another 20 bright, isolated stars in the field, for aperture sizes from 1 to 
20 pixels, in increments of 0.05 pixels.  
Of the 20 field stars, we selected as comparisons the four that produced 
the most stable flux ratios with respect to the target, i.e.~the differential 
light curves with the least photometric dispersion during the predicted 
out--of--eclipse phases.  
The locations of these four best comparisons are shown along with the target 
in Figure~\ref{finder}.  
A fixed aperture size of 6.1 pixels (1.67 arcsec) and a 14--24 pixel (3.82--6.55 
arcsec) sky annulus produced the most stable differential photometry when 
combining the curves from all four comparisons.  
The average photometric dispersion in the out--of--eclipse portion of the 
combined light curve was 0.781\%, a factor of 3.02 larger than the expected 
noise per datapoint.

However, the combined curve also showed clear systematic trends that could be 
attributed to atmospheric effects such as seeing and airmass, as well as 
instrumental effects such as changes in the (x,y) position of the stars in 
the images or temperature and pressure changes in the instrument throughout 
the duration of the observations.

These trends were individually investigated by fitting a linear correlation between 
each parameter and the differential flux, for the out--of--eclipse points only.  
The most significant systematic trends correlated with variations in the seeing, 
and we found that nearly all the systematic noise in the light curves was removed 
after removing a trend based on that parameter.

For the final photometry we de-trended each differential light curve with respect 
to seeing, and then combined the four, ending up with a dispersion of 0.547\%, 
2.11 times the expected noise limit.  The final light curve from this analysis 
is shown in the top panel of Figure \ref{lightcurve}.
  
\subsection{Analysis B}\label{subsec:raa_b}

The second analysis approach, analogous to that of \citet{2009A&A...493L..31S},
began with the separation of the two dither position sets into different light 
curves, and implemented the SysRem algorithm \citep{2005MNRAS.356.1466T} for 
de-correlation.

We used standard IDL procedures to perform aperture photometry on 19 stars, 
including the target, for aperture sizes between 1.0 and 14.9 pixels spaced in 
increments of 0.1 pixel.  The centers of each star were determined by fitting a 
2D moffat function to the point spread function (PSF) of each star.  
The residual sky background was determined by finding the sky annulus which 
resulted in the highest target signal--to--noise (photometry error with photon, 
sky, and readnoise).  
On the basis of this evaluation, the best sky annulus resulted to be one 
of inner radius 19 pixels and outer radius 20 pixels, 
corresponding to 120 pixels between 5.19 and 5.46 arcsec from the star.

The five stars selected as comparisons in this analysis are shown in 
Figure~\ref{finder}.  
The comparison selection criterium was the same as in Analysis A, i.e.~the field 
stars that produced the most stable photometry with respect to the target during 
the out--of--eclipse phase.  
Later comparison showed that the four reference stars used in Analysis A overlap 
with these five.  

After producing individual light curves for each dither position, we 
de-correlated the curves using an implementation of the SysRem algorithm 
\citep{2005MNRAS.356.1466T}, which seeks to minimize the expression 

\begin{equation} \sum(r_{ij}-c_ia_j)^2/\sigma^2_{ij}, \label{eq:sysrem}\end{equation}

\noindent where $r_{ij}$ is the average-subtracted stellar magnitude for the $i$th 
star of the $j$th image, $\sigma$ is the uncertainty of $r_{ij}$, $c_i$ is an 
epoch--dependent parameter, and $a_j$ is a stellar--dependent parameter.  
The first pass through the SysRem algorithm produced a light curve with a 
single clear linear trend.  
We found this linear trend to be efficiently removed by either a second pass 
through the SysRem algorithm or by allowing the baseline flux to vary in time 
linearly, fit by two parameters.  
Removing this linear trend effectively removed any detectable systematic trends.  
We fit each dither position individually at first, but found that the linear 
slope was similar for both, and re--combined the sets.

In order to test the effectiveness of our de-correlation procedures, we searched 
for the presence of residual systematic errors correlated in time (``red--noise'', 
\citealt{2006MNRAS.373..231P}) by checking that the binned residuals followed an 
$N^{-1/2}$ relation, when binning in time by $N$ points.  
The presence of red-noise causes the variance to follow a 
$\sigma^2=\sigma_w^2/N+\sigma_r^2$ relation, where $\sigma_w$ is the uncorrelated 
white noise component while $\sigma_r$ characterizes the red-noise.  
We found no significant evidence for red-noise when binning on time scales up to 
42 minutes (100 points), showing that any correlated noise had been effectively 
removed by the de-trending. 

The optimal aperture size of 5.6 pixels (1.53 arcsec) was determined to be that 
which minimized the standard deviation of the light curve when using the target 
and reference stars in these de-trending procedures.  
The final light curve from this analysis had a dispersion of 0.661\%, 2.55 times 
the expected noise limit.  The final light curve from this analysis is 
shown in the second panel of Figure \ref{lightcurve}.

\subsection{Secondary Eclipse Fit}\label{subsec:raa_eclipsefit}

The next step in our analysis was to search for the eclipse signal from the 
planet in the observed light curves.  
This was done separately for each light curve resulting from analyses A and B, 
and the solutions were then combined to produce a final result.

We used the orbital period from \citet{2009arXiv0903.1845B} 
and the stellar and planetary radius from \citet{2009arXiv0905.4571G}
(see Table \ref{table2}) to generate a grid of eclipse models with no 
limb darkening and a constant out-of-eclipse baseline.  The free 
parameters in the model grid were the baseline level and the depth and 
central phase of the eclipse, with the grid covering baseline levels 
between 99.6 and 100.4\% in increments of 0.002\%, 
eclipse depths between 0 and 0.8\% in 0.002\% steps, and central phases 
between 0.46 and 0.54 in 0.0001 phase increment steps.  
Running each light curve through the model grid resulted in the following 
best--fit solutions: 
For the analysis A light curve, a baseline of 100.014\% and an eclipse 
of depth 0.322\%, centered at phase 0.5003; for the analysis B light curve, 
a baseline of 100.006\%, an eclipse depth of 0.355\%, and a central phase 
of 0.5073.  
We decided to use the point--by--point average of the light curves from 
analyses A and B for our final result, shown in the third panel of 
Figure \ref{lightcurve}.  
Table \ref{fullflc} shows the first few unbinned points in the final result, 
and the full dataset is available in the online version.  
The best-fitting model in that case, found via chi-square minimization 
and with a reduced $\chi^{2}$ of 1.051, gave a baseline of 
100.010$^{+0.042}_{-0.040}$\%, an eclipse depth of 0.336$^{+0.068}_{-0.064}$
and a central phase 0.5022$^{+0.0023}_{-0.0027}$.  
These values are given, along with the other parameters of the system, in 
Table \ref{table2}.  
The results of each fit for the entire parameter space 
are shown in Figure \ref{depthphasecontour}. 

To estimate the error in the eclipse depth, we put all the in-eclipse 
points from the de-trended photometry into a single bin, the entire 
out-of-eclipse portion into a second bin, and combined the binned errors.  
We had 295 in-eclipse points with an average individual dispersion 
of 0.531\%, and 344 out-of-eclipse points with an average individual 
dispersion of 0.528\%, which combined to give us a 1$\sigma$ error of 0.042\%.  
For the error in phase we have directly adopted the 1$\sigma$ confidence values 
from the contour plot in Figure \ref{depthphasecontour}.

As an additional test to confirm the depth of the eclipse, we generated histograms 
of the distribution of normalized flux for both the in-eclipse and out-of-eclipse 
portions of the light curve, adopting as central eclipse phase the 
value $\psi$ = 0.5022 obtained above. The result, shown in Figure \ref{histogram},
shows a clear 0.336\% shift of the distribution of in-eclipse points with respect 
to the out-of-eclipse points, in full agreement with the result of the model 
grid fits.

\section{Atmosphere Model Fits}\label{sec:models}

While secondary eclipse detections in a single broad band, such as the optical 
detections with the CoRoT data \citep{2009Natur.459..543S,2009arXiv0906.2814A}, 
cannot constrain a planet's thermal and reflected light independently of one 
another, we can begin to disentangle the reflected and thermal contributions 
to the total light from the planet by comparing the secondary eclipse depths 
at different wavelengths.  
Combining our Ks-band eclipse detection of \corot with the other three 
detections by \citet{2009Natur.459..543S}, \citet{2009arXiv0905.4571G}, and 
\citet{2009A&A...501L..23A}, in this section we make the first simultaneous 
multi--wavelength comparison of observations of an exoplanet atmosphere to 
current models at optical and near--infrared wavelengths.

In the following subsections we compare the observations first to simple 
blackbody models and then to more sophisticated radiative--convective models 
of irradiated planetary atmospheres in chemical equilibrium.  
The advantage of the blackbody models is that they provide simple initial 
estimates ofthe global properties of the planetary atmospheres, such as their 
approximate temperatures, 
their reflective properties, and how efficiently they redistribute energy from the 
irradiated to the non-irradiated sides (see e.g.~\citealt{2007ApJ...667L.191L}).  
Atmosphere models provide more details about specific properties of the 
atmospheres, such as finer spectral information.  
This allows the identification of specific absorbing or emitting chemicals, 
as well as their specific atmospheric depths, and the presence of thermal 
inversion layers (see e.g.~\citealt{2008ApJ...678.1436B}).

\subsection{From Contrasts to Flux Densities}\label{subsec:models_fluxdensities}

Before beginning the model comparisons, we converted the measured
eclipse depths to planetary fluxes, because secondary eclipse photometry
does not provide an absolute measure of the planet's brightness, but a
planet--to--star flux ratio integrated through the instrument and
filter profile.

For this conversion, we calculated the synthetic flux density of a Kurucz (1993) 
G0V--type star, equivalent to the CoRoT-1b host star, in the CoRoT White, 
CoRoT Red, NB2090, and Ks filters using each filter's response curve.  
The CoRoT White, NB2090, and Ks filters have well-defined transmission 
functions centered at 0.6, 2.09, and 2.15 $\mu$m, respectively, but the response 
curve for the CoRoT Red channel is not as well-known.  The collected light is 
passed through a prism that divides it into red, green, and blue portions, but 
the behavior is different for each star.  
Following \citet{2009Natur.459..543S}, we estimated a wavelength cutoff of 
$\sim$560 nm by taking the fraction of light in the red channel to the total 
light, giving an effective wavelength of 710 nm.  
To approximate the behavior of the red channel, we used the same response curve 
as for the white channel, but with zero transmission below 560 nm.

For the purposes of the comparison we used a hypothetical distance of 10 pc 
(i.e.~the physical flux density of the star and planet were derived for this 
standard scale distance, not their actual distance).  
We then multiplied each synthetic stellar flux density by the observed 
planet--to--star contrast in that band.  
The resulting planet flux densities are summarized in Table~\ref{table1}, 
together with the observed contrasts and the parameters of each filter.  
The derived planet flux densities are also shown in Figure~\ref{C1}.

We note that the contrasts observed and the flux densities calculated
for \corot in the NB2090 and the Ks filters are very similar, our measurement 
confirming the detection made by \citet{2009arXiv0905.4571G}.

\subsection{Blackbody--based Planet Models}\label{subsec:models_blackbody}

In the blackbody approximation, the temperature profile of an irradiated planet 
will be a smooth combination of reflected and thermally emitted light.  
That temperature profile is determined by the interplay of the stellar 
irradiation, Bond albedo $A_B$, and a re-radiation factor $f$, which describes 
how efficiently energy from incident radiation is transported around the planet 
before being re-emitted.  Following \citet{2007ApJ...667L.191L}, the 
value $f$=2/3 corresponds to no redistribution before re-radiation and $f$=1/4 
corresponds to the incident energy being evenly redistributed around the planet.  
The combination of the optical and near--infrared measurements of \corot 
provide powerful constraints on both the reflected light (i.e.~planetary albedo) 
and the heat redistribution between the day and night sides of the planet.  
In this section we apply models combining reflected light and thermal blackbody 
radiation to interpret the detections. The results are illustrated in  Figures 
\ref{C1} and \ref{D1}.

Figure~\ref{C1} includes the four measured planetary flux densities, 
normalized to a distance of 10 pc, along with models with different temperature, 
albedo and re-radiation factor values.  
The effect of increasing the albedo is to increase the flux density measured in 
the optical (which corresponds to reflected light), while decreasing the infrared 
emission due to the lower absorbed energy.  
Increasing the day--night side heat redistribution lowers the entire spectral 
energy distribution, but does not shift the balance between the optical and 
near--infrared flux densities. The models shown in the figure reveal that 
the observations are entirely inconsistent with efficient day--night heat 
redistribution and also exclude a high albedo. 

Figure~\ref{D1} shows the result of exploring the entire A$_B$ -- $f$ parameter 
space using the observed flux densities and the thermal emission plus reflected 
light models.  
To quantify the constraints and identify the best fit, we evaluated these models 
along a grid spanned by Bond albedos A$_B$=0 to 0.3 and re-radiation factors 
f=1/4 to 2/3, calculating at each grid point the predicted temperature, the 
predicted flux densities in the {\bf four} observed bands, and a $\chi^2$ value.  
The contour levels in the figure show the total four--dimensional distance 
of each model from the observations, expressed in units of uncertainties.  
This was calculated by summing the squares of the differences between the 
observed and predicted flux densities divided by the uncertainty at each 
wavelength, and taking the sum's square root, i.e.~equivalent to the square 
root of the $\chi^2$ value.  
The best fit corresponds to a model with dayside temperature 
2454$^{+84}_{-170}$~K, a very low albedo (A$_B$=0.000$^{+0.087}_{-0.000}$), 
and inefficient but measurable heat redistribution ($f$=0.450$^{+0.065}_{-0.085}$).  
These parameters are given in the third section of Table~\ref{table2}.  
Thus, our simple blackbody--based modeling suggests a low--albedo planet with 
some measurable levels of energy re-distribution, but still a very prominent 
temperature difference between the day and night sides.

\subsection{Theoretical Atmosphere Models}\label{sec:models_TheoreticalModels}

Due to prominent molecular absorption bands heavily--irradiated giant planet 
atmospheres are thought to display an emission spectrum very unlike a blackbody.  
Thus, while the blackbody models provide useful first estimates of the planetary 
properties, realistic atmospheric models are required to derive the actual 
physical properties of the planets.  
To gain more detailed information about the physical and chemical processes 
undergoing in the atmosphere of CoRoT-1b, we compared the observations to the 
latest, and still evolving, models of irradiated hot Jupiter atmospheres.

The model atmospheres we used are derived from self-consistent coupled radiative 
transfer and chemical equilibrium calculations, based on the models described 
in \citet{2000ApJ...538..885S,2003ApJ...588.1121S}, \citet{2003ApJ...594.1011H}, 
and \citet{2005ApJ...625L.135B,2006ApJ...650.1140B,2008ApJ...678.1436B}.  
The most important components of the code include molecular and atomic opacities, 
and calculations to determine the chemical abundances using thermochemical models 
(e.g.~\citealt{2007ApJS..168..140S}).  
The day and night sides of the planet are treated separately, with the 
day side receiving incident flux from the star using the appropriate 
\citet{1993yCat.6039....0K} spectral model, and the night side receiving heat 
from the day side via convection.  
The convection is modeled with a mixing length equal to the pressure scale height 
\citep{2008ApJ...678.1436B}.  
The heat redistribution, described by a parameter P$_n$, is used to derive the 
planet's flux at the time of secondary eclipse \citep{2008ApJ...682.1277B}.  
The P$_n$ parameter represents the fraction of the incident stellar 
energy that is redistributed to the night side (P$_n$=0 corresponds to no 
redistribution; P$_n$=0.5 is uniform distribution around the planet).
\footnote{
P$_n$=0 (no energy redistribution) 
corresponds to $f$=2/3 and P$_n$=0.5 (maximum redistribution) 
corresponds to $f$=1/4.  However, as the physical models incorporate factors 
that are accounted for differently than in the blackbody models (e.g.~pressure, 
opacity), the P$_n$-$f$ relation is degenerate, i.e.~multiple P$_n$ values may 
correspond to the same $f$ value.
}

The immediate conclusion from applying the base models described above is that 
atmospheres with no thermal inversion, unable to reproduce the near--infrared 
fluxes we observe, can be confidently excluded.  
Therefore we modified those models by adding an extra optical (0.37 to 1.0 
$\mu$m) absorber with a constant opacity $\kappa_e$ to the abundances predicted 
by the chemical equilibrium calculations.  
This absorber, placed at an atmospheric height (i.e.~pressure) of P = 
0.01 bar, has the effect of creating a strong temperature inversion, with 
the extra optical absorption heating the stratosphere.  
Some of the resulting representative models are shown in Figure~\ref{E1}.

The upper panel in the figure shows identically irradiated planets, but with 
different redistribution parameters (P$_n$ = 0.1, 0.3, 0.5) and extra absorber 
opacities ($\kappa_e$ = 0.0, 0.1, 0.0 cm$^2$ g$^{-1}$).  
Models with no energy redistribution (i.e.~P$_n$=0.0) were too bright in 
the optical regime to fit the two CoRoT points.  
Although representing diverse planetary atmospheres, none of these three, nor any 
of the other models, reproduced the observed very bright near-infrared flux densities.

In an attempt to more closely reproduce the high observed near-infrared fluxes, we 
used an atmosphere model with P$_n$=0.1 and $\kappa_e$ = 0.05 cm$^2$ g$^{-1}$.  
The extra absorber had to be placed deeper in the atmosphere (at $\sim$0.1 bar).  
This is a factor of 10--100 deeper than what has been used to fit the Spitzer data 
for other hot jupiters, in which case an extra optical absorber 
at $\sim$10$^{-3}$ -- 10$^{-2}$ bar reproduced the IRAC points 
(e.g.~\citealt{2008ApJ...684.1427M}).  
The temperature at this layer, also calculated by the model, is around 2200 K.
The result of this final model attempt is illustrated in the lower panel of 
Figure~\ref{E1}.  
This model fits the NB2090 and CoRoT channel observations fairly well (within 
0.9 $\sigma$), but still underpredicts the observed Ks-band flux density by 2.6 $\sigma$.  
Still, there is some improvement over the upper--panel models, for which the Ks--band 
deviations are between 3.3 and 4.9 $\sigma$.  
The parameters in this best--fit model are also listed in Table \ref{table2}.

Previous to this work, there were virtually no observational constraints 
on hot Jupiters in this wavelength range.  
Although a perfect model match to the near--infrared data has yet to be achieved, the 
results reveal very hot temperatures at low--to--moderate optical depths, i.e.~in the 
upper stratospheres probed by the 2-micron observations (e.g.~Figure 1 in 
\citealt{2008ApJ...678.1436B}).

\section{Discussion}\label{sec:discussion}

The detections of exoplanets via transits and secondary eclipses have produced many of 
the most valuable insights to their physical properties, but also yielded some puzzling 
results.  {Among these is the apparent division of the population of hot Jupiters 
into two distinct classes based on their atmospheric properties: one group that appears 
cooler, with water and methane absorption bands and more efficient energy redistribution, 
and another with higher levels of thermal emission, strong day--night contrasts, and 
lacking the expected absorption bands.
The dominant explanation for this dichotomy is the presence in many planets of a thermal 
inversion layer, i.e.~a hot stratosphere caused by extra absorbers of optical light 
\citep{2003ApJ...594.1011H,2008ApJ...682.1277B,2008ApJ...678.1419F}.  

The exact nature of the absorbers is a difficult question still being investigated.  
\citet{2009arXiv0902.3995S} recently argued that vanadium oxide is not likely to fulfill 
this role, and that the previously favored titanium oxide would require unusually high 
levels of macroscopic mixing to remain in the upper atmosphere.  S$_2$, S$_3$, and HS 
compounds, as absorbers of optical and ultraviolet light, have also recently been both 
considered and questioned as causes of the thermal inversion 
\citep{2009AAS...21430601Z,2009arXiv0903.1663Z}.  
Consistent modeling of the infrared secondary eclipse spectra of six planets by 
\citet{2008ApJ...678.1436B} suggest that the presence of the necessary absorber may 
depend not only on the incident stellar radiation, but also on planetary 
metallicity and surface gravity.

From the combination of the {\bf four} optical to near--infrared detections 
we have detailed in this paper, it appears that \corot clearly falls into the class of 
hot Jupiters with a thermal inversion.
The best--fit physical model to the combined dataset
requires a small redistribution parameter P$_n$ = 0.1 and an extra optical absorber 
with flat opacity $\kappa$ = 0.05 cm$^2$/g.  
While we cannot determine the identity of this absorber, the model constrains its 
altitude, requiring absorption a factor of 10--100 times deeper in the atmosphere than 
suggested by previous results for other exoplanets.  
This new constraint also highlights the power of combined optical and near-infrared 
photometry.

\corot reflects some other surprising trends as well.  
Hot Jupiters tend to have very low albedos 
\citep{2000ApJ...538..885S,2006ApJ...646.1241R,2008ApJ...682.1277B}, and our findings 
place this planet's atmosphere in agreement with that trend.  
From the best--fit physical model derived in Section \ref{sec:models_TheoreticalModels}, 
we obtained an estimate of the planet's geometric albedo at the wavelengths probed by 
the CoRoT detections (0.4 to 1.0$\mu$m) to be A$_g$ = 0.05$\pm$0.01.  
Assuming a wavelength-independent Lambert sphere (see \citealt{2007ApJ...667L.191L}), 
this corresponds to a Bond albedo A$_B$=0.075$\pm$0.015.

\corot is also notable for its extremely large radius (1.45 R$_{Jup}$, 
\citealt{2009arXiv0905.4571G}); evolutionary models for a hot Jupiter of its mass, age, 
and irradiation predict a radius of only 0.94 - 1.18 R$_{Jup}$ 
\citep{2007ApJ...659.1661F}.  
This places it towards the upper end of the wide distribution of radii that has been 
seen among the population of transiting exoplanets.  

While small--radius planets can be modeled with a larger, denser core, the inflated 
sizes of planets like \corot provide quite a challenge to planetary models.  
Several explanations have been proposed, including heat retained by enhanced 
atmospheric opacities \citep{2007ApJ...668L.171B}, deposition of kinetic wind energy 
in the upper atmosphere \citep{2008ApJ...682..559S}, and significant tidal 
heating caused by orbital eccentricity or a rapidly-rotating star 
\citep{2003ApJ...592..555B,2009ApJ...698L..42G,2009AAS...21430602M,2009arXiv0902.3998I}.  
Though \citet{2009arXiv0905.4571G} measured an eccentricity $e$=0.071$^{+0.042}_{-0.028}$ 
for CoRoT--1b's orbit, the eccentricity we measure 
($e\cos{\omega}$=0.0035$^{+0.0036}_{-0.0042}$) based on its mid--eclipse phase is 
consistent within its errors with a circular orbit.   
Thus, there is no evidence that tidal heating derived from a currently eccentric orbit 
contributes to the energy budget of CoRoT--1b, although intense heating in the recent 
past cannot be excluded (e.g.~\citet{2009ApJ...702.1413M}).  
In each of the diverse models, the radius of the planet is determined by the
pressure--temperature structure of the atmosphere and the balance between the 
energy input and output.  
In order to understand and improve the models, we must first constrain the energy 
balance by understanding the basic atmospheric properties, such as chemical abundances, 
pressure--temperature profiles, and energy redistribution.

With our detection in Ks joining the 0.6, 0.71 and 2.09 $\mu$m measurements, we 
are beginning to get a more detailed picture of the atmospheric properties of CoRoT--1b.  
However, there are still sizeable gaps in the observed spectrum, and even 
the most advanced current models have difficulty explaining the high flux 
levels in the 2-$\mu$m window.  
It is crucial to add further detections in both narrow and broad bands, in the optical 
and near-infrared, and a larger target sample to provide stronger constraints to the 
planetary atmosphere models.

\section{Summary}\label{sec:summary}

-- We directly detect thermal emission from CoRoT--1b in Ks-band, determining the 
planet-to-star flux ratio to be 0.336$\pm$0.042\%.  

-- Using simple blackbody--based models, we find the best fit to be a blackbody of 
2454$^{+84}_{-170}$~K, and confidently rule out both a high albedo and efficient 
day--night heat redistribution.

-- Using realistic atmosphere models, we find the need for a thermal inversion layer 
and an extra absorber near the 0.1--bar level, deeper in the atmosphere than the 
Spitzer mid--infrared data suggested for other hot jupiters.

-- Both the blackbody models and physical atmospheric models agree on a small 
but non-zero amount of heat redistribution, and a Bond albedo less than 0.09.

In short, the combined optical to near--infrared photometry of \corot has allowed 
us to independently constrain the reflected light and thermal emission, and has 
revealed a very hot, low--albedo planet with a large day--night contrast and a 
prominent temperature inversion.

\acknowledgments{

J.C.R. and D.A. are grateful for the critical support provided 
from the Space Telescope Science Institute 
Director's Discretionary Research Fund D0101.90131.  
M.L-M. acknowledges support provided by NASA through Hubble Fellowship
grant HF-01210.01-A awarded by the STScI, which is operated by the
AURA, Inc. for NASA, under contract NAS5-26555. D.K.S. is supported by
CNES. A.B. is supported in part by NASA grant NNX07AG80G. This work
has been partially supported by the National Science Foundation
through grant AST--0908278.

{\it Facilities:} \facility{APO (NICFPS)}

}



\clearpage

\appendix



\begin{table}[ht]
\begin{center}
\begin{tabular}{lccccc}
  \hline
  Filter & $\lambda_{eff}$ ($\mu$m) & FWHM ($\mu$m) & Peak Trans. & Planet:Star Flux Ratio & Pl. Flux Density (Jy)\\
  \hline
  CoRoT white & 0.60 & 0.42 & 72\% & 0.016$\pm$ 0.006\%$^1$ & 0.0093$\pm$0.0035\\
  CoRoT red & 0.71 & 0.25 & 72\% & 0.0126$\pm$ 0.0033\%$^2$ & 0.0087$\pm$0.0023\\ 
  NB 2090 & 2.095 & 0.020 & 82\% & 0.278$^{+0.043}_{-0.066}$\%$^3$ & 0.1094$^{+0.0169}_{-0.0260}$\\
  Ks & 2.147 & 0.318 & 97.5\% & 0.336$\pm 0.042$\%$^4$ & 0.1172$\pm$0.0160\\
  \hline
\end{tabular}
\caption{Transmission information of each detection filter, along with the planet-to-star 
flux ratio and physical flux density of the planet in each filter.  The physical flux 
densities are calculated for a hypothetical distance of 10 pc.  {\bf References.} 
(1) \citealt{2009A&A...501L..23A}, 
(2) \citealt{2009Natur.459..543S}, 
(3) \citealt{2009arXiv0905.4571G}, 
(4) this work.
\label{table1}
}
\end{center}
\end{table}

\begin{table}[ht]
\begin{center}
\begin{tabular}{cccc}
  \hline
  HJD & Orbital Phase & Observed Flux & Flux Error \\
  \hline
     2,454,846.686842 & 0.4320984781 & 0.9939594865 & 0.0052779813 \\
     2,454,846.687166 & 0.4322515726 & 1.0044958591 & 0.0052779813 \\
     2,454,846.687398 & 0.4324662685 & 1.0090936422 & 0.0052779813 \\
     2,454,846.687722 & 0.4326120615 & 0.9946241379 & 0.0052779813 \\
     2,454,846.687942 & 0.4328347445 & 1.0004521608 & 0.0052779813 \\
  \hline
\end{tabular}
\caption{Final light curve, calculated by averaging the point-by-point average of the results from analyses A and B.  Note: Table 2 is published in its entirety in the electronic edition of the {\em Astrophysical Journal}.  A portion is shown here for guidance regarding its form and content.
\label{fullflc}
}
\end{center}
\end{table}

\begin{table}[ht]
\begin{center}
\begin{tabular}{lccccc}
  \hline
  Stellar Parameter & Value & Unit & Ref.\\
  \hline
  Stellar mass $M_\star$ & 1.01$^{+0.13}_{-0.22}$ & $M_\sun$ & 3\\
  Stellar radius $R_\star$ & 1.057$^{+0.055}_{-0.094}$ & $R_\sun$ & 3\\
  Star T$_{eff}$ & 5950$\pm$150 & K & 1\\
  log $g$ & 4.25$\pm$0.30 & (cgs) & 1\\
  $[M/H]$ & -0.3$\pm$0.25 & dex & 1\\
  Right ascension (J2000) & 06 48 19 & & 1\\
  Declination (J2000) & -03 06 08 & & 1\\
  \hline
  Measured Planet Parameter & Value & Unit & Ref \\
  \hline
  Planet mass $M_p$ & 1.07$^{+0.13}_{-0.18}$ & $M_J$ & 3\\
  Planet radius $R_p$ & 1.45$^{+0.07}_{-0.13}$ & $R_J$ & 3\\
  Transit epoch $T_0$ & 2454159.452879$\pm$0.000068 & HJD & 2\\
  Orbital period & 1.5089686$^{+0.0000005}_{-0.0000006}$ & day & 2 \\
  Semi-major axis $a$ & 0.0259$^{+0.0011}_{-0.0020}$ & AU & 3\\
  Orbital inclination $i$ & 85.66$^{+0.62}_{-0.48}$ & degrees & 3\\
  Mid-eclipse phase & 0.5022$^{+0.0023}_{-0.0027}$ & & 4\\
  Orbital eccentricity $e\cos{\omega}$ & 0.0035$^{+0.0036}_{-0.0042}$ & & 4\\
  2.2 $\mu$m eclipse depth & 0.336$\pm$ 0.042\% & & 4\\
  \hline
  Blackbody model-derived Planet Parameter & Value & Unit & Ref \\
  \hline
  Bond albedo $A_B$ & 0.000$^{+0.087}_{-0.000}$ & & 4\\
  Re-radiation factor $f$ & 0.450$^{+0.065}_{-0.085}$ & & 4\\
  Blackbody model temperature T$_{bb}$ & 2454$^{+84}_{-170}$ & K & 4\\
  \hline
  Physical model-derived Planet Parameter & Value & Unit & Ref\\
  \hline
  Bond albedo $A_B$ & 0.075$\pm$0.015 & & 4\\
  Energy redistribution factor P$_n$ & 0.1 & & 4\\
  Absorber opacity $\kappa_e$ & 0.05 & cm$^2$g$^{-1}$ & 4\\
  Absorber depth in atmosphere & 0.1 & bar & 4\\
  Temperature at absorber depth & 2200 & K & 4\\
  \hline
\end{tabular}
\caption{Star and planet parameters.  
{\bf References.} 
(1) \citealt{2008A&A...482L..17B}, 
(2) \citealt{2009arXiv0903.1845B}, 
(3) \citealt{2009arXiv0905.4571G}, 
(4) this work. 
\label{table2}
}
\end{center}
\end{table}


\begin{figure*}
 {\centering
  \includegraphics[width=0.58 \textwidth,angle=90]{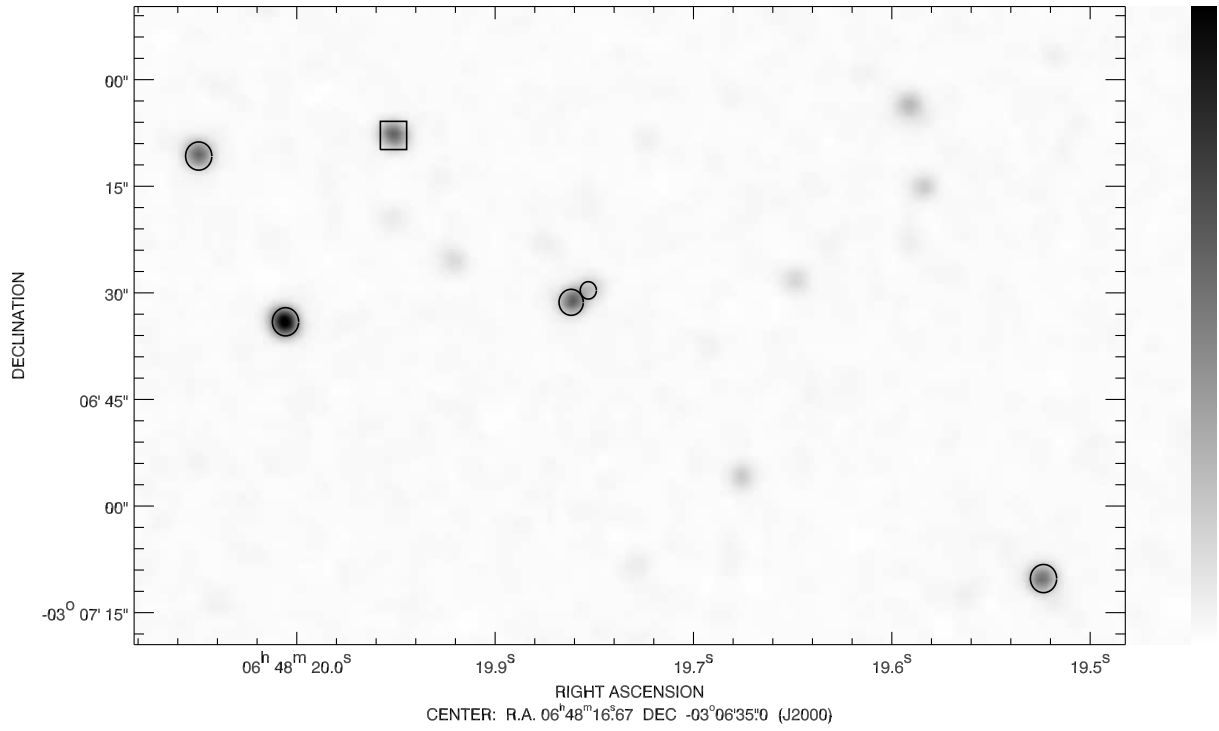}}
\caption[]{Finder chart for CoRoT--1 target (enclosed in square) and the field stars selected 
as photometric comparisons.  The four stars used in Analysis A are enclosed in the larger 
circles, while Analysis B used those four plus the star in the smaller circle.
\label{finder}
}
\end{figure*}

\begin{figure*}
 {\centering
  \includegraphics[width=1.00 \textwidth,angle=0]{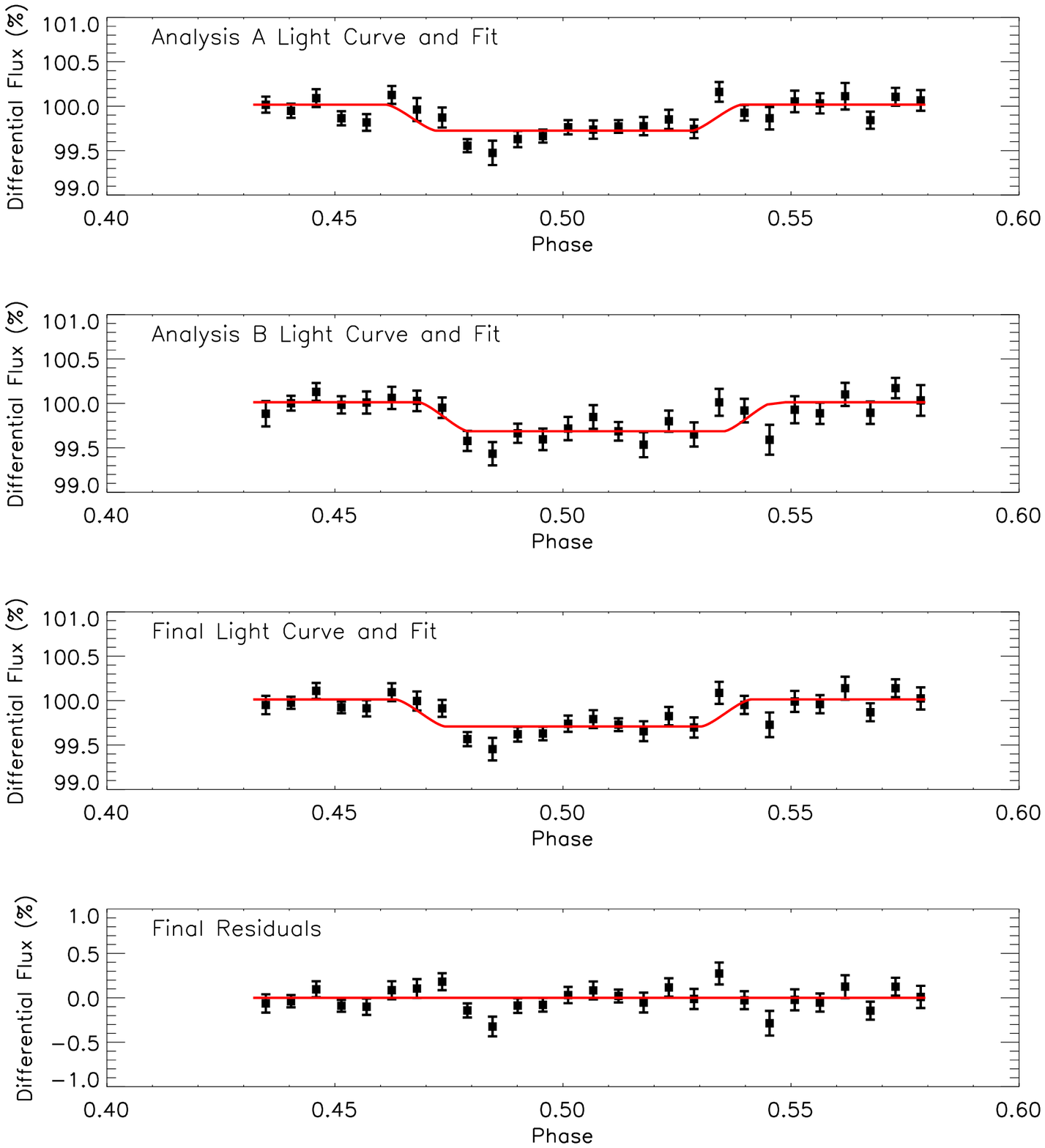}}
\caption[]{
{\it Top Panel}: Light curve from Analysis A after de-trending, 
with its best-fit model.

{\it Second Panel}: Light curve and best-fit model from Analysis B after 
de-trending and removal of the remaining linear trend.

{\it Third Panel}: Final light curve calculated by combining Analyses A and B.  
A model with the best-fit secondary eclipse is 
shown as the horizontal red line, with  best-fit central phase shift of 0.5022.  
The baseline is set at 100.010\%, with an eclipse depth of 0.336\%.

{\it Bottom Panel:} Flux residuals (Observed - Model) for the combined light curve. 

In all plots, each point corresponds to a 12 minute bin.  

\label{lightcurve}
}
\end{figure*}

\begin{figure*}
 {\centering
  \includegraphics[width=0.6 \textwidth,angle=90]{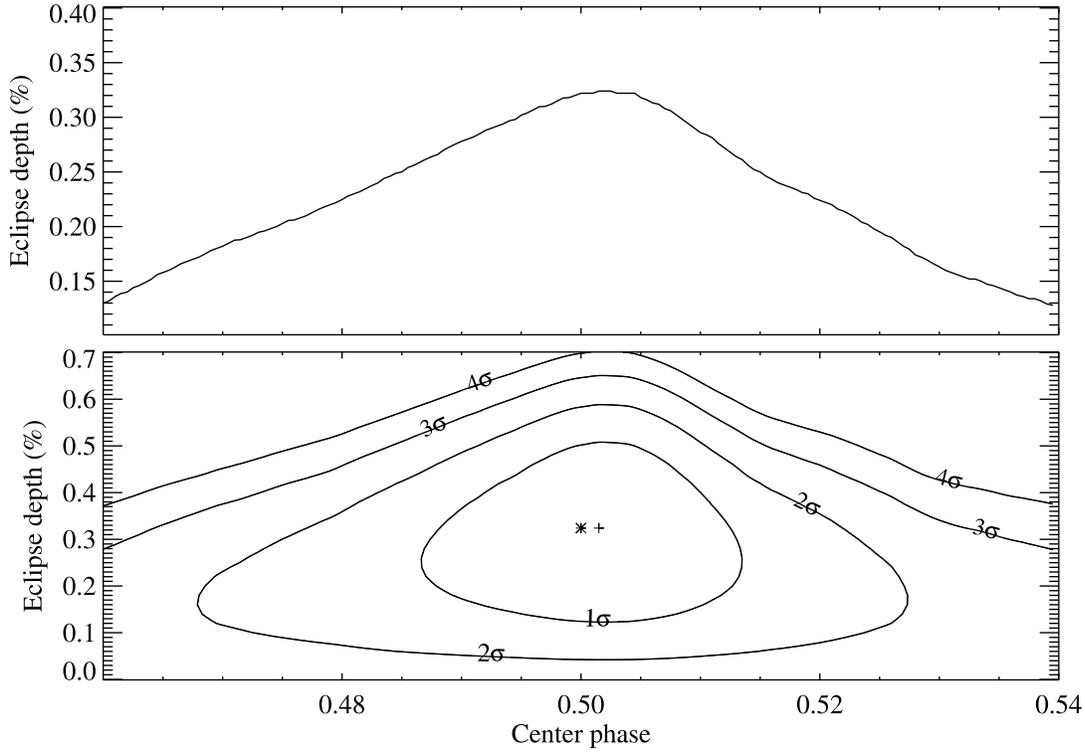}}
\caption[]{
{\it Top Panel}: Model eclipse depth versus central phase for 
phases between phases 0.46 and 0.54.
The best model fit to the data has a depth of 0.336$\%$ and central phase 
0.5022. The eclipse depth falls rapidly in both directions away from that phase value.
{\it Bottom Panel}: Confidence contours of the best fit at the 68.3$\%$, 95.5$\%$ 
and 99.7$\%$ level.
The best-fit value is indicated as a {\bf cross} 
at phase 0.5022. The star at 0.5 
indicates the phase of the expected center of the eclipse for a circular orbit. 
The result is therefore consistent with a zero eccentricity orbit 
(e $\cos{\omega}$= 0.0035 $^{+0.0036}_{-0.0042}$).
\label{depthphasecontour}
}
\end{figure*}

\begin{figure*}
 {\centering
  \includegraphics[width=1.\textwidth,angle=0]{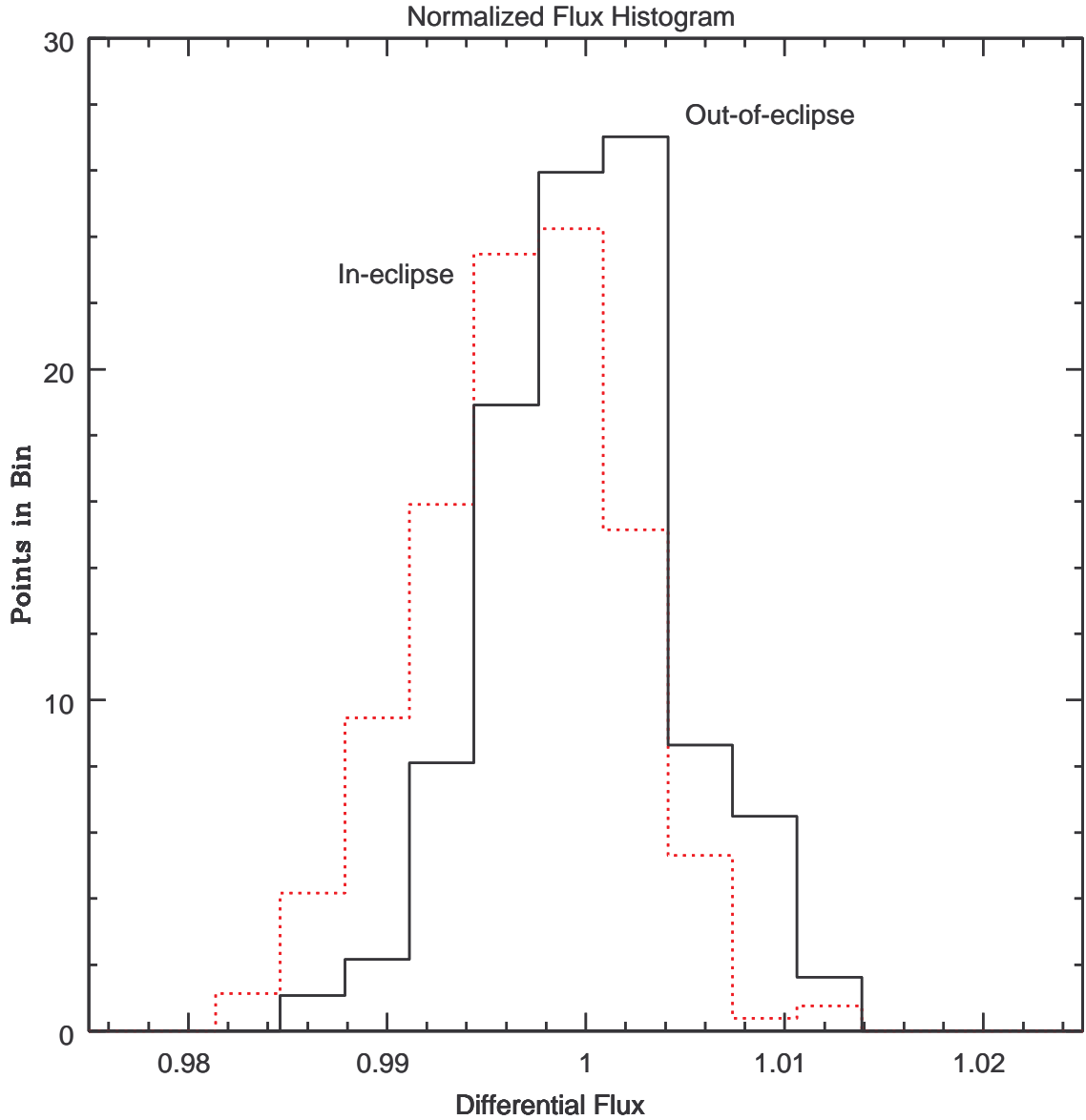}}
\caption[]{Normalized flux histograms of the in-eclipse (red dotted line) 
and out-of-eclipse (black solid line) portions of the \corot light curve 
in Figure \ref{lightcurve}. 
The width of each bin is 0.336\%, the same as the
detected eclipse depth.
\label{histogram}
}
\end{figure*}

\begin{figure*}
 {\centering
  \includegraphics[width=1.\textwidth,angle=0]{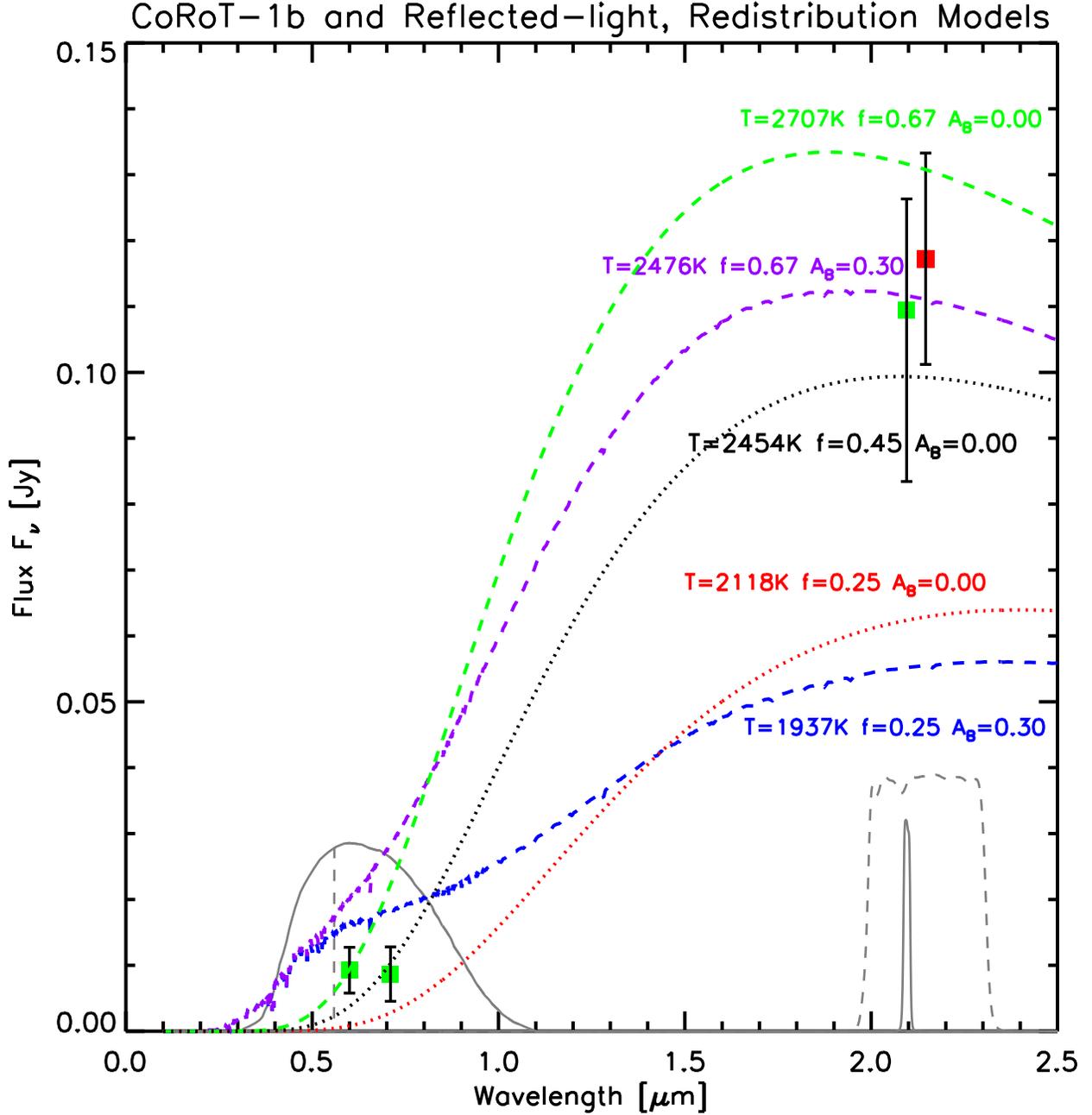}}
\caption[]{Comparison of the four detected \corot planet fluxes with models that 
include reflected light and heat redistribution, for a range of Bond albedo (A$_B$) 
and re-radiation factors ($f$).  
We show two models with maximum energy redistribution ($f$ = 1/4) and 
Bond albedos of 0 (red) and 0.3 (blue), and 
two models with no energy redistribution ($f$ = 2/3) and Bond albedos of 0 (green) 
and 0.3 (purple).  
The best-fit model, shown in black, is for a zero-albedo planet with a small, 
but non-zero amount of heat redistribution.

At the bottom of the figure we show scaled transmission functions for each of the 
filters used: CoRoT the dashed line shows the blue cutoff for the red channel), 
NB2090 (solid), and Ks (dashed).
}
\label{C1}
\end{figure*}

\begin{figure*}
 {\centering
  \includegraphics[width=1.\textwidth,angle=0]{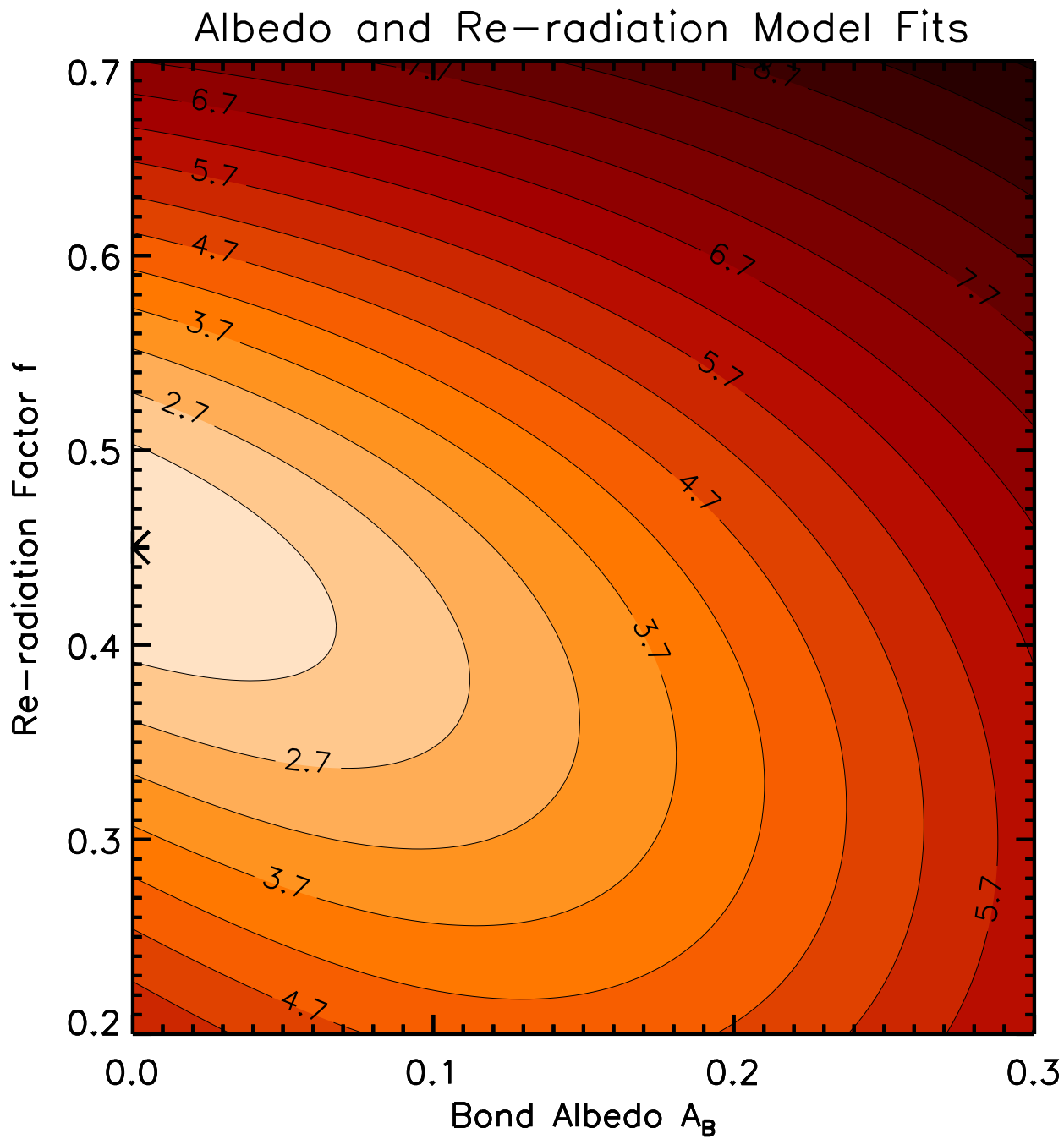}}
\caption[]{Contour plot showing the best-fit albedo and re-radiation factors for 
the planet fluxes detected.  
The contour levels show the total four--dimensional distance 
of each model from the observations, expressed in units of uncertainties.  
This was calculated by summing the squares of the differences between the 
observed and predicted flux densities divided by the uncertainty at each 
wavelength, and taking the sum's square root, i.e.~equivalent to the square 
root of the $\chi^2$ value.
The data clearly favor a very low-albedo planet with 
{\bf inefficient but measurable} energy redistribution, 
with the best fit (indicated by a bold X) at A$_B$=0.000$^{+0.087}_{-0.000}$, 
$f$=0.450$^{+0.065}_{-0.085}$, producing a temperature of 2454$^{+84}_{-170}$~K.
\label{D1}
}
\end{figure*}

\begin{figure*}
 {\centering
  \includegraphics[width=1.\textwidth,angle=0]{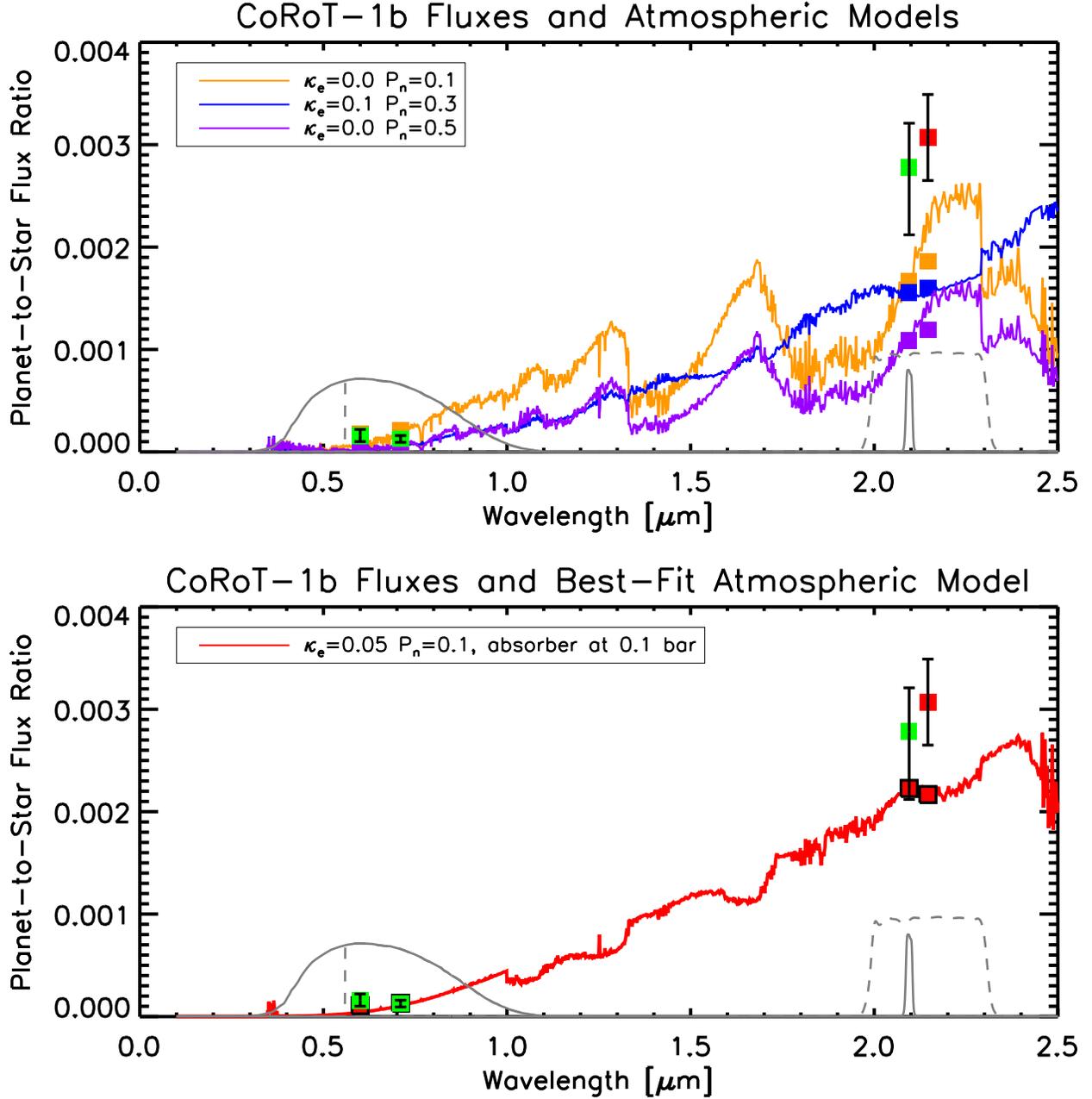}}
\caption[]{
{\it Top Panel}: The measured planet-to-star flux ratios compared to the band--averaged 
ratios from atmospheric models that incorporate extra optical absorbers placed near the 
0.01 bar level.  
Three models shown here in orange, blue, and purple, have absorber opacities 
$\kappa_e$= 0.0, 0.1, and 0.0 cm$^2$ g$^{-1}$, and redistribution parameters 
P$_n$ = 0.1, 0.3, and 0.5, respectively.  

{\it Bottom Panel}: The measured flux ratios compared to the predicted ratios from the 
best--fit atmospheric model, with $\kappa$=0.05 cm$^2$ g$^{-1}$ and P$_n$ = 0.1, and the 
absorber placed near the 0.1 bar level, deeper in the atmosphere than for the other models.  
\label{E1}
}
\end{figure*}

\end{document}